\documentclass[%
 reprint,
 amsmath,amssymb,
 aps,
]{revtex4-2}

\usepackage{graphicx}
\usepackage{dcolumn}
\usepackage{bm}
\usepackage{xcolor}
\usepackage{amsmath}
\usepackage{comment}
\usepackage{enumitem}

\newcommand{\ed}{\end{document}}
\newcommand{\beq}{\begin{equation}}
\newcommand{\eeq}{\end{equation}}
\newcommand{\beqa}{\begin{eqnarray}}
\newcommand{\eeqa}{\end{eqnarray}}
\newcommand{\bc}{\begin{center}}
\newcommand{\ec}{\end{center}}

\newcommand{\ba}{\begin{array}}
\newcommand{\ea}{\end{array}}
\newcommand{\pa}{\partial}

\begin{document}

\preprint{APS/123-QED}

\title{Acoustic Kerr Metric in  Analogue Gravity
}

\author{Deeshani Mitra}
\email{deeshani1997@gmail.com}
\author{Surojit Dalui}
\email{surojitdalui003@gmail.com}
\author{Subir Ghosh}%
 \email{subirghosh20@gmail.com}
\affiliation{Indian Statistical Institute, \\ 203, Barrackpore Trunk Road, Kolkata 700108, India}
\author{Arpan Krishna Mitra}
\email{arpankmitra@aries.res.in}
 \affiliation{Aryabhatta Research Institute of Observational Sciences (ARIES),\\ Manora Peak Nainital - 263001, Uttarakhand, India. }

\date{\today}

\begin{abstract}
\noindent	
The present paper is based on a previous work (involving two of the present authors) where a generalized fluid dynamical model was proposed. The underlying symplectic structure of the Lagrangian discrete degrees of freedom obeyed a Non-Commutative algebra, generated by Berry curvature correction. In an Euler (or Hamiltonian) framework, this is manifested as an extended algebra between the fluid variables, leading to the extended fluid model. Here we study the dynamics of sonic fluctuations  that live in this effective analogue gravity  spacetime. Interestingly enough, the effective metric  resembles  that of a spinning Black Hole; the spin is induced by the underlying Non-Commutative structure. The effective mass and spin parameters of the Black Hole, in terms of fluid parameters, are also identified. The connection of our model with anomalous Hall systems may lead to observable signatures of the analogue black hole in physical systems.  
\end{abstract}

\maketitle


\section{\label{sec:level1}Introduction:
}
\noindent
In a seminal paper \cite{unru},  Unruh was the first to explicitly show that in an irrotational, non-viscous, barotropic flow, linearisation  of the perturbed  fluid equations over a background flow (known solution of basic  hydrodynamic equations, {\it{i.e.}} continuity and Euler equations) results in a (structurally relativistic) wave equation of the velocity potential in an effectively curved spacetime. This naturally led to the similarity between the propagation of sound waves in a supersonic flow and  the propagation of scalar fields in a black hole space-time with the sound velocity in the former creating a horizon. The corresponding wave equation in fluid yields the acoustic metric, totally dictated by the background fluid parameters - in other words, Analogue Gravity (an exhaustive  review is provided in  \cite{rev}. This is conceptually akin to a   massless scalar field’s propagation  in a background space-time geometry in General Relativity. 
A (by no means exhaustive) list of Analogue Gravity examples in diverse systems is provided in \cite{analog}.

In recent times there has been a surge of activity in trying to find different physical systems that can sustain analogue gravity scenarios that can be obtained in a laboratory setup. (For recent activity and references, see \cite{recent}.) We, in the present work, pursue a different route in revealing a new analogue gravity model; in particular in an extended fluid model approach. Two of the present authors in a  previous paper \cite{mitra2021} have   developed a generalized   fluid model in a Hamiltonian framework, generating an extended set of  continuity and Eulerian equations. The modified  Poisson bracket structure of the Euler variables, (density and velocity fields), exploited above, resulted from an effective noncommutative particle algebra with Berry curvature  effects, that constitute the discrete fluid degrees of freedom in  the Lagrangian approach. 
 Incidentally, this phase space characterizes the  semi-classical  electron dynamics in a magnetic Bloch band, in the presence of a   periodic potential with an external  magnetic field and Berry curvature \cite{duval}. Berry curvature and induced magnetic field in momentum space are responsible for the "anomalous velocity" in quantum Hall effect \cite{nagaosa}. The generalized fluid dynamics can be relevant in electron hydrodynamics in condensed matter, {\it{i.e.}} in situations where electron flow is influenced by  hydrodynamic laws instead of being fully Ohmic  \cite{das}. In normal circumstances,   electrons in metals behave as a nearly-free Fermi gas since the effective mean free path for electron-electron collision is quite large, allowing  impurities and lattice thermal vibrations (phonons) to destroy a  collective viscous  fluid-like  electron motion. But in recent years,  the hydrodynamic regime has been achieved   in extremely pure,   high quality, electronic materials - especially graphene \cite{graf}, layered materials with very high electrical conductivity such as metallic delafossites  $PdCoO_2  ,~ PtCoO_2$ \cite{met}. 
 
In the present work, we have developed an  analogue gravity picture in the extended fluid model, essentially following similar mathematical steps as \cite{unru}. In order to derive the acoustic metric we have perturbed the extended ("anomalous" since it manifests a divergence anomaly \cite{mitra2021} ) fluid equations. 

Now comes the most interesting and novel part of our work. First of all, we show that, under certain restrictions, the acoustic metric, after a coordinate transformation (see for example \cite{Natario2009}, bears a close similarity with the Kerr metric \cite{kerr}, written in Eddington-Finkelstein (EF) coordinates \cite{Dalui2022}. Although a fluid analogue of Kerr metric has not been reported earlier to the best of our knowledge, it is probably not surprising in our system since it has been demonstrated long ago by one of us \cite{sg} that a spin-like feature appears in Berry curvature modified particle dynamics.

The other novelty in our analysis is the following: instead of trudging the conventional path of looking at the  dynamics in this analogue gravity background, we perform a deeper study of the analogy between Kerr and our acoustic metric. Using only the fluid parameters and performing a dimensional analysis, we explicitly construct the  fluid composites  that are identified with effective mass, $m_{eff}$, and angular momentum per unit mass $a_{eff}$. It is very intriguing to note that this somewhat heuristic matching of parameters survives throughout the metrics; that is $m_{eff}$ and  $a_{eff}$ in acoustic metric appear in similar positions as $m$ and $a$ in Kerr metric. 
 
The paper is organized as follows: In Section II we derive the extended form of  acoustic metric starting from   Berry curvature corrected fluid dynamics. In Section III, a coordinate transformation is introduced to rewrite the acoustic metric to facilitate a  comparison between the acoustic metric and the Kerr metric (for spinning massive object in General Relativity). Subsequently exploiting the structural similarity, we construct the  analogue mass and spin/unit mass parameter of the acoustic metric. Section IV is devoted to concluding remarks and future open problems.

\section{The  acoustic metric with Berry curvature effects}
\noindent
Berry curvature corrected fluid dynamic equations, namely the continuity and Euler equations, as derived in  \cite{mitra2021} are respectively

\begin{equation}
	\label{con}
	\dot{\rho}+{\bf{\nabla}}\cdot{\bf J}^{an}=e\rho {\bf{\mathcal{F}}}\cdot({\bf{\nabla}}\times {\bf E}) 
	\end{equation}
where the anomalous current ${\bf J}^{an}$ is
\begin{equation}
	\label{J}
{\bf J}^{an}=\left(\frac{\rho{\bf{v}}}{\mathcal{A}}\right)+ e \rho({\bf{\mathcal{F}}}\cdot{\bf{v}}){\bf{B}}+e \rho ({\bf E}\times {\bf{\mathcal{F}}})+\mathcal{F}\times \nabla P 
\end{equation}
and
\begin{widetext}
\begin{eqnarray}
\dot{{\bf{v}}}+ \frac{({\bf{v}}\cdot{\bf{\nabla}}){\bf{v}}}{\mathcal{A}} = && -\frac{{\bf{\nabla}}P}{\rho\mathcal{A}} +e\frac{\rho {\bf{v}}\times{\bf{B}}}{\rho\mathcal{A}}-e \frac{{\bf{B}\cdot{\bf{\nabla}}P}}{\rho}{\bf{\mathcal{F}}}-e({\bf{v}}\cdot{\bf{\mathcal{F}}})
({\bf{B}}\cdot{\bf{\nabla}}){\bf{v}}+\Bigg[\left(\frac{\nabla P}{\rho}\times \mathcal{F}\right)\cdot\nabla
-\frac{1}{\rho}\nabla v^2 \cdot\{\nabla \times  {\bf({\mathcal{F}\rho)}}\}\nonumber\\
&&+ 2 v^2 \left({\bf{\mathcal{F}}}\times\frac{\nabla \rho}{\rho}\right)\cdot\nabla - {\bf{\mathcal{F}}}\cdot \left(\frac{\nabla \rho}{\rho}\times \nabla v^2\right)\Bigg]{\bf{v}}-e\Bigg[ \frac{{\bf{E}}}{\mathcal{A}}+ e ({\bf{E}}\cdot{\bf{B}}){\bf{\mathcal{F}}}-({\bf{E}}\times{\bf{\mathcal{F}}})\cdot {\bf{\nabla}}{\bf{v}}\Bigg] \label{eulber}
\end{eqnarray}	
\end{widetext}	
with,
\begin{eqnarray}
\mathcal{F}_{i}({\bf x,\bf k})=\frac{\Omega_{i}}{1+e{\bf{B}}({\bf x})\cdot{\bf{\Omega}(\bf k)} }, 
\end{eqnarray}
\begin{eqnarray}
\mathcal{A}({\bf x,\bf k})=1+e{\bf{B}}({\bf x})\cdot{\bf{\Omega}(\bf k)}~.
\end{eqnarray}
We consider a barotropic fluid with $P(\rho)$ being the pressure. $e$ represents the electronic charge, ${{\bf{E}},\bf{B}}$ denote respectively  the external electric and magnetic field and ${\bf{\Omega_{i}}}$ is the $i$'th component of ${\bf{\Omega}}({\bf {k}})=\nabla_{{\bf {k}}}\times {\bf {A}}({\bf {k}})$, the Berry curvature  in momentum $(\Bar{k})$ space.

Note that in \cite{mitra2021}, we had shown that the fluid model \eqref{con} and \eqref{eulber}, possess an Adler-Bell-Jackiw type divergence anomaly $\sim e^2{{\bf{E}}.\bf{B}}$ and it would be interesting to consider the effect of anomaly in the analogue gravity context but for computational simplicity, in the present work we restrict to ${\bf{E}}=0$. Furthermore, we  
assume  the Berry curvature ${\bf\Omega(k)} $ is small which in turn makes $\mathcal{F}$ negligible whenever multiplied by any first order term. Hence we  now proceed with a reduced  form of the fluid equations.
\begin{equation}
\label{bcont}
     \dot\rho=-\nabla\left(\frac{\rho \bf{v}}{\mathcal{A}}\right),
\end{equation}
\begin{equation}
\label{beul}
    \dot{{\bf{v}}}+ \frac{({\bf{v}}\cdot{\bf{\nabla}}){\bf{v}}}{\mathcal{A}}=-\frac{{\bf{\nabla}}P}{\rho\mathcal{A}}~. 
\end{equation}
Considering ${\bf{E}}$ to be irrotational, we introduce a velocity potential $\psi$ as,
\begin{equation}
    \label{valpo}
    \bf{v}=-\nabla \psi~.
\end{equation}
The velocity $c_s$ of sonic disturbance in the medium and the system enthalpy $h$ is defined in the following way,
\begin{equation}
c_s=\sqrt{\frac{dP}{d\rho}}, ~~
\nabla h=\nabla P/\rho~
\label{nab}
\end{equation}
and \eqref{beul} is now rewritten as;
\begin{equation}
\label{valpo2}
     -\nabla{\dot{\psi}}+\nabla\left[\frac{(\nabla{\psi})^{2}}{2\mathcal{A}}\right]=-\nabla{\frac{h}{\mathcal{A}}}    
\end{equation}
thereby yielding
\begin{equation}
\label{valpo1}
   -\dot\psi+\frac{(\nabla{\psi})^{2}}{2\mathcal{A}}=-{\frac{h}{\mathcal{A}}}~.
\end{equation}
The signature of the Berry curvature correction in the much-simplified system manifests in the factor ${\mathcal{A}}$.
 Following  \cite{unru} we express the fluid variables as a background and first-order perturbation term, 
\begin{equation}
    \nonumber
   \rho =\rho_0+\epsilon \rho_1, ~P=P_0+\epsilon P_1,~v_{i}=v_{0i}+\epsilon v_{1i}=\partial_i\psi_0+\epsilon\partial_i\psi_1 ~. 
\end{equation}
From \eqref{nab} we find,
\begin{equation}
    \label{1}
   P=P_0+\epsilon c_s^2\rho_1,~
\nabla h_1=c_s^2\nabla \rho_1/\rho_0~.
\end{equation}
The first order perturbation terms of \eqref{valpo1} with \eqref{1} give 
 \begin{equation}
    {\rho}_{1}=\left(\frac{\rho_{0}\dot{\psi_{1}}}{c_{s}^{2}}\right)+\left(\frac{\rho_{0} \vec{v_{0}} \cdot \nabla{\psi_{1}}}{c_{s}^{2}\mathcal{A}}\right)~.
    \label{nabol}
\end{equation}
On the other hand, the first-order contribution  of \eqref{con} is
\begin{eqnarray}
\label{hibus}
    \dot{\rho}_{1}= -\frac{1}{\mathcal{\mathcal{A}}}\nabla .\left( \rho_{1}\vec{v_{0}}-\rho_{0}\nabla \psi_{1}\right)~.
\end{eqnarray}
To compare \eqref{nabol} and \eqref{hibus} we take time derivative of \eqref{nabol}
\begin{equation}
    \dot {\rho}_{1}=\partial_{t}\left(\frac{\rho_{0}\dot{\psi_{1}}}{c_{s}^{2}}\right)+\partial_{t}\left(\frac{\rho_{0} \vec{v_{0}} \cdot \nabla{\psi_{1}}}{c_{s}^{2}\mathcal{A}}\right)
    \label{nabolt}~.
\end{equation}
After some non-trivial algebraic steps involving \eqref{hibus} and \eqref{nabolt} we arrive at the wave equation, bearing resemblance with a massless scalar field equation in a curved spacetime:
 \begin{equation}
     \label{effwav}
     \pa_{\mu}(f^{\mu\nu}\pa_{\nu}\psi_{1})=0~.
 \end{equation}

Matrix representation of $f^{\mu\nu}$ is given below
$$f^{\mu\nu}= \frac{\rho_{0}}{c_s^{2}}\left(
\begin{array}{cccc}
 \mathcal{A} & v_x & v_y & v_z \\
 v_x & \frac{v_x^2-c_s^2}{\mathcal{A}} & \frac{v_x v _y}{\mathcal{A}} & \frac{v_x v_z}{\mathcal{A}} \\
 v_y & \frac{v_x v_y}{\mathcal{A}} & \frac{v_y^2-c_s^2}{\mathcal{A}} & \frac{v_y v_z}{\mathcal{A}} \\
 v_z & \frac{v_x v_z}{\mathcal{A}} & \frac{v_y v_z}{\mathcal{A}} & \frac{v_z^2-c_s^2}{\mathcal{A}} \\
\end{array}
\right)$$
Note that, the metric depends on the background (or zero'th order) velocity ${v}_{0i}$ and for notational simplicity, from now on we  use $v_i$ in place of ${v}_{0i}$. 
To Obtain the expression of the effective background metric, we define $g^{\mu\nu}$ 
\begin{equation}
\label{gandf}
    f^{\mu\nu}=\sqrt{-g}~g^{\mu\nu}~
\end{equation}
with the determinant of $f^{\mu\nu}$ given by
\begin{equation}
    \vert{f^{\mu\nu}}\vert={(\sqrt{-g})}^{4}\frac{1}{g}= g=-\frac{\rho_{0}^{4}}{c_{s}^{2}\mathcal{A}^{2}}~.
\end{equation}
With these results in hand, we can invert $g^{\mu\nu}$ to get the cherished effective metric structure $g_{\mu\nu}$ leading to the line element $ds^{2}$ 
\begin{equation}
ds^{2} = g_{tt} dt^2+2 g_{ti}  dtdx^{i}+g_{ij} d x^{i} d x^{j}
\end{equation}
 with
 \begin{eqnarray}
g_{\mu\nu}= \frac{\rho_{0}}{\mathcal{A} c_s}\left(
\begin{array}{cccc}
 \frac{c_s^2-({v_x}^2+{v_y}^2+{v_z}^2)}{\mathcal{A} } & v_x & v_y & v_z \\
 v_x & -\mathcal{A} & 0 & 0 \\
 v_y & 0 & -\mathcal{A} & 0 \\
 v_z & 0 & 0 & -\mathcal{A} \\
\end{array}
\right)\nonumber
 \end{eqnarray}
This is the  effective background metric structure over which the first-order sonic disturbance in a charged fluid, subjected to Berry curvature correction, will propagate. This constitutes our main result. There are now two parallel channels for further investigation: the study of the dynamics of wave motion in this non-trivial background and looking for experimentally measurable signatures and analyzing the physics behind this metric with possible identification of analogue effective parameters of the fluid. In the present work, we pursue the second alternative.

For subsequent work, we express the analogue metric in polar form,
 \begin{eqnarray}
g_{\mu\nu}= \frac{\rho_{0}}{\mathcal{A} c_s}\left(
\begin{array}{cccc}
 \frac{c_s^2-(v_{r}^2+v_{\theta}^{2}+v_{\phi}^{2})}{\mathcal{A} } & v_{r}& r v_{\theta} & r\sin{\theta}v_{\phi} \\
 v_{r} & -\mathcal{A} & 0 & 0 \\
 r v_{\theta} & 0 & -\mathcal{A}r^{2} & 0 \\
 r\sin{\theta}v_{\phi} & 0 & 0 & -\mathcal{A}r^{2}\sin^{2}{\theta} \\
\end{array}
\right)\nonumber
 \end{eqnarray}
We adopt a scenario where the background fluid flow is  spherically symmetric so that  $v_{\theta}=v_{\phi} =0$ with only  the radial velocity component  $v_r$ surviving.
\section{Analogue metric of the Kerr type  }
\noindent
Till date, the most popular form of  the Analogue metric is of Schwarzschild nature, but we now show that indeed, the analogue metric derived here, even with the above-mentioned approximations, bears a richer structure, possibly of the Kerr form, the metric that describes a spinning black hole in GR \cite{kerr}.

Incorporating these simplifying assumptions, we end up with the acoustic path, 
\begin{widetext}
\begin{eqnarray}
	ds_{\text{ac}}^{2}=\frac{\rho_{0}}{\mathcal{A}c_s}\left[\frac{\left(c_s^{2}-v_{r}^{2}\right)}{\mathcal{A}} dt^{2} + 2v_{r} dt dr - \mathcal{A}\{dr^{2} + r^{2}d\theta^{2} + r^{2}\sin^{2}\theta d\phi^{2}\}\right]~.\label{acoustic_metric_spherically_sym}
\end{eqnarray}
\end{widetext}
In order to facilitate a meaningful identification of dimensional parameters of our acoustic metric with a metric in GR, we follow two steps. In the first step, we convert the acoustic path length dimension to  $|ds^{2}|=(\text{length})^{2}=[L]^{2}$. In GR, the metrics can comprise dimensional parameters such as Newtonian gravitational constant $G$ and velocity of light $c$, among others. In a similar way, the Acoustic metric can depend on $c_s$ (not a constant in general), background fluid density $\rho_0$ (not a constant in general) etc.  This scheme requires us to introduce another conventional fluid parameter known as the dynamic (or absolute) viscosity $\mu$ of dimension of $|\mu|=[M][L]^{-1}[T]^{-1}$ (with the kinematic viscosity  being $ \mu/\rho_{0}$). We also introduce a length scale $l$ (which can be the spatial dimension of the fluid) into our acoustic system. This yields the acoustic metric (\ref{acoustic_metric_spherically_sym})   
\begin{widetext}
	\begin{eqnarray}
	ds_{\text{ac}}^{2}=\frac{c_s l\rho_{0}}{\mu\mathcal{A}}\left[\frac{\left(c_s^{2}-v_{r}^{2}\right)}{\mathcal{A}} dt^{2} + 2v_{r} dt dr - \mathcal{A}\{dr^{2} + r^{2}d\theta^{2} + r^{2}\sin^{2}\theta d\phi^{2}\}\right]~.\label{acoustic_metric_spherically_sym_with param}
	\end{eqnarray}
\end{widetext}
such that $|ds_{\text{ac}}^{2}|=(\text{length})^2$.

In the second step,  we perform a coordinate transformation 
\begin{widetext}
\begin{eqnarray}
	dt\rightarrow dt+\frac{dr}{c_s} + \frac{d\theta}{\omega_s} + \frac{ d\phi}{\Omega_s}.\label{generalised_transformation}
\end{eqnarray}
\end{widetext}
where $c_s$, $\omega_s$ and $\Omega_s$ are the sound velocity, the angular frequency and the azimuthal frequency of the sonic disturbance. The Acoustic metric is transformed to 
\begin{widetext}
	\begin{eqnarray}
	ds^{2}_{\text{ac}}=&&\frac{c_s l\rho_{0}}{\mu\mathcal{A}}\Bigg[\frac{\left(c_s^{2}-v_{r}^{2}\right)}{\mathcal{A}} dt^{2}   +  2\left\{\frac{(c_s^{2}-v_{r}^{2})}{c_s\mathcal{A}} + v_{r}\right\} dtdr +   2\left\{ \frac{(c_s^{2}-v_{r}^{2})}{\omega_s\mathcal{A}}\right\} dtd\theta +  2\left\{ \frac{(c_s^{2}-v_{r}^{2})}{\Omega_s\mathcal{A}}\right\} dtd\phi \nonumber\\
	&& + \left\{\frac{\left(c_s^{2}-v_{r}^{2}\right)}{c_s^{2}\mathcal{A}} + 2\frac{v_{r}}{c_s} - \mathcal{A}  \right\} dr^{2}  + \frac{2}{\omega_s}\left\{\frac{\left(c_s^{2}-v_{r}^{2}\right)}{c_s \mathcal{A}} + v_{r}  \right\}drd\theta  +  \frac{2}{\Omega_s}\left\{\frac{\left(c_s^{2}-v_{r}^{2}\right)}{c_s \mathcal{A}} +  v_{r}  \right\}drd\phi\nonumber\\
	&& +  \left\{ \frac{(c_s^{2}-v_{r}^{2})}{\omega_s^{2}\mathcal{A}} -\mathcal{A} r^{2} \right\}d\theta^{2} + 
	2\left\{ \frac{(c_s^{2}-v_{r}^{2})} {\omega_s\Omega_s\mathcal{A}}\right\}d\theta d\phi + \left\{ \frac{(c_s^{2}-v_{r}^{2})}{\Omega_s^{2}\mathcal{A}} - \mathcal{A}r^{2}\sin^{2}\theta \right\}d\phi^{2}\Bigg]~.\label{acoustic_metric_with_ac_param}
	\end{eqnarray} 
\end{widetext}
We now perform the final task: the construction  of analogue fluid parameters to be identified with a suitable GR metric. We start by writing down the Kerr metric that represents a stationary, axisymmetric black hole  in Eddington-Finkelstein (EF) coordinates (with the metric signature (+,-,-,-)) \cite{Dalui2022}
\begin{widetext}
\begin{eqnarray}
ds^{2}_{\text{Kerr}}=&&\Bigg(1-\frac{2Gmr}{c^2\Sigma^{2}}\Bigg)c^2dt^{2}-\frac{4Gmr}{c\Sigma^{2}}dtdr+\frac{4Gmra}{c^2\Sigma^{2}}\sin ^{2} \theta dt d\phi-\Bigg(1+\frac{2Gmr}{c^2\Sigma^{2}}\Bigg)dr^{2}+2\frac{a}{c}\sin^{2}\theta\Bigg(1+\frac{2Gmr}{c^2\Sigma^{2}}\Bigg)drd\phi\nonumber\\
&&-\Sigma^{2}d\theta^{2}-\Bigg(r^{2}+\frac{a^{2}}{c^2}+\frac{2Gmra^{2}\sin^{2} \theta}{c^4\Sigma^{2}}\Bigg)\sin^{2}\theta d\phi^{2}~. \label{Kerr_metric}
\end{eqnarray}
\end{widetext}
where $m$ is the mass of the Kerr black hole, $a=J/m$ is the angular momentum per unit mass and
\newline
$\Sigma^{2}=r^{2}+a^{2}\cos^{2}\theta$. 

Now comes the  interesting part. The Acoustic (\ref{acoustic_metric_with_ac_param}) and Kerr (\ref{Kerr_metric}) metrics are structurally similar on the $\theta=\pi/2$ hyper-surface, having the explicit forms shown below:
\begin{widetext}
	\begin{eqnarray}
	ds^{2}_{\text{ac}}=&&\frac{c_s l\rho_{0}}{\mu\mathcal{A}}\Bigg[\frac{\left(c_s^{2}-v_{r}^{2}\right)}{\mathcal{A}} dt^{2}   +  2\left\{\frac{(c_s^{2}-v_{r}^{2})}{c_s\mathcal{A}} + v_{r}\right\} dtdr +  2\left\{ \frac{(c_s^{2}-v_{r}^{2})}{\Omega_s\mathcal{A}}\right\} dtd\phi + \left\{\frac{\left(c_s^{2}-v_{r}^{2}\right)}{c_s^{2}\mathcal{A}} + 2\frac{v_{r}}{c_s} - \mathcal{A}  \right\} dr^{2} \nonumber\\
	&&    +  \frac{2}{\Omega_s}\left\{\frac{\left(c_s^{2}-v_{r}^{2}\right)}{c_s \mathcal{A}} +  v_{r}  \right\}drd\phi + \left\{ \frac{(c_s^{2}-v_{r}^{2})}{\Omega_s^{2}\mathcal{A}} - \mathcal{A}r^{2} \right\}d\phi^{2}\Bigg]~,\label{acoustic_metric_with_ac_param_at_pi/2}
	\end{eqnarray} 
\end{widetext}
\begin{widetext}
	\begin{eqnarray}
	ds^{2}_{\text{Kerr}}=&&\Bigg(1-\frac{2Gm}{r c^2}\Bigg)c^2dt^{2}-\frac{4Gm}{r c}dtdr+\frac{4Gma}{r c^2} dt d\phi-\Bigg(1+\frac{2Gm}{r c^2}\Bigg)dr^{2}+2\frac{a}{c}\Bigg(1+\frac{2Gm}{r c^2}\Bigg)drd\phi\nonumber\\
	&&-\Bigg(r^{2}+\frac{a^{2}}{c^2}-\frac{2Gma^{2}}{r c^4}\Bigg) d\phi^{2}~. \label{Kerr_metric_at_pi/2}~
	\end{eqnarray}
\end{widetext}
We can now exploit the dimensional equality $ds^{2}_{\text{ac}}=ds^{2}_{\text{Kerr}}=(\text{length})^2$ to construct the analogue mass and spin parameters for Acoustic metric, using only the fluid parameters introduced here.
\begin{itemize}
	\item Comparing the dimensions of $g_{tt}$  we get
	\begin{eqnarray}
	\text{dim}\Bigg\vert\Bigg(1-\frac{2Gm}{r c^2}\Bigg)c^2\Bigg\vert =\text{dim}\Bigg\vert \frac{c_s l\rho_{0}}{\mu\mathcal{A}^{2}}\Bigg[\left(1-\frac{v_{r}^{2}}{c_s^{2}}\right)c_s^{2}\Bigg]\Bigg\vert~.\label{dt2_term}
	\end{eqnarray}
so that 
\begin{eqnarray}
\text{dim} |m_{\text{eff}}|\equiv \text{dim} \Bigg\vert\left(\frac{r c^{2}}{G} \right) \left( \frac{c_s l \rho_{0} v_{r}^{2}}{\mu \mathcal{A}^{2} c_{s}^{2}}\right)\Bigg\vert~.\label{mass_eff}
\end{eqnarray}
Dimension of $r c^{2}/G$ is $[L][L^{2} T^{-2}]/[L^{3} T^{-2} M^{-1}]$, so that
\begin{equation}
    \text{dim}\Bigg\vert\frac{rc^{2}}{G}\Bigg\vert 
=\text{dim}\Bigg\vert \frac{\mu l^{2}}{c_s}\Bigg\vert~.
\label{gtt}
\end{equation}
Therefore, from (\ref{mass_eff}) we can write down the effective mass parameter of the acoustic metric in terms of the fluid parameters,
\begin{eqnarray}
m_{\text{eff}}\equiv \frac{l^{3}\rho_{0}v_r^{2}}{\mathcal{A}^{2} c_s^{2}}\label{mass_eff_final}
\end{eqnarray}
	\item Comparing the dimensions of $g_{t\phi }$  we get 
	\begin{eqnarray}
	\text{dim}\Bigg\vert	\frac{4Gma}{r c^2} \Bigg\vert =  \text{dim}\Bigg\vert \frac{c_s l\rho_{0}}{\mu\mathcal{A}^{2}}   \left\{ 2 \frac{(c_s^{2}-v_{r}^{2})}{\Omega_s\mathcal{A}}\right\}\Bigg\vert~.\label{dtdphi_term}
	\end{eqnarray}
Thus the effective rotation parameter can be written as, (ignoring the 1/2 factor as this is dimensionless),
\begin{eqnarray}
	\text{dim} |a_{\text{eff}}|\equiv \text{dim} \Bigg\vert\left(\frac{r c^{2}}{G m}\right)\left\{ \frac{c_s l\rho_{0}}{\mu\mathcal{A}^{2}} \right\}  \left[ \frac{(c_s^{2}-v_{r}^{2})}{\Omega_s\mathcal{A}}\right]\Bigg\vert~.
\end{eqnarray}	
Therefore, replacing $m$ by $m_{eff}$ from  (\ref{mass_eff_final}) the final expression of the analogue rotation parameter for  fluid is
\begin{eqnarray}
a_{\text{eff}}\equiv \frac{l \rho_{0} c_s^{5}}{\mu \mathcal{A}^{2}\Omega_s v_{r}^{2}}~.
\end{eqnarray}
\end{itemize} 
In general, the effective parameters can depend on position but as special cases, they can be treated as constants.
Finally, in terms of the effective mass parameter ($m_{\text{eff}}$) and the effective rotation parameter ($a_{\text{eff}}$) the acoustic metric in (\ref{acoustic_metric_with_ac_param_at_pi/2}) is written as
\begin{widetext}
\begin{eqnarray}
	ds_{\text{ac}}^{2} = && \left(\frac{c_s l \rho_{0}}{\mu\mathcal{A}^{2}} - m_{\text{eff}}\frac{c_s}{\mu l^{2}}\right)c_s^{2}dt^{2} + 2 \frac{m_{\text{eff}}}{\mu l^{2}}\left(c_s^{2} - v_{r}^{2} + c_s v_{r} \right)dt dr + 2 m_{\text{eff}} a_{\text{eff}} \left(\frac{\mu\mathcal{A}v_{r}^{2}}{c_s^{5} l \rho_{0}}\right)\left(c_s^{2} - v_{r}^{2}\right) dt d\phi\nonumber\\
	&&+\left[\frac{c_s l \rho_{0}}{\mu\mathcal{A}}\left(1+2\frac{v_r}{c_s}-\mathcal{A}\right)-m_{\text{eff}}\frac{c_s}{\mu l^{2}} \right]dr^{2} + 2 a_{\text{eff}} \frac{v_{r}^{2}}{c_s^{2}}\left[\frac{1}{c_{s}\mathcal{A}}\left\{1-m_{\text{eff}}\frac{\mathcal{A}^{2}}{l^{3}\rho_{0}}\right\}+\frac{v_{r}}{c_s^{2}} \right]dr d\phi\nonumber\\
	&&+\left[\frac{\mu v_{r}^{4}}{c_s^{7} l \rho_{0}}a_{\text{eff}}^{2}\left(1-m_{\text{eff}}\frac{\mathcal{A}^{2}}{l^{3}\rho_{0}}\right)  -\frac{c_s l \rho_{0}}{\mu}r^{2}\right]d\phi^{2}~.\label{final_metric_in_terms_m_a}
\end{eqnarray}
\end{widetext}
Introducing a dimensionless parameter, $\sigma (r)=v_r(r)/c_s(r)$ we find
\begin{equation}
    \label{sig}
  m_{\text{eff}}\equiv \frac{l^{3}\rho_{0}\sigma^2}{\mathcal{A}^{2}} , ~~ a_{\text{eff}}\equiv \frac{l \rho_{0} c_s^{3}\sigma^2}{\mu \mathcal{A}^{2}\Omega_s } 
\end{equation}
and rewrite
\begin{widetext}
\begin{eqnarray}
	ds_{\text{ac}}^{2} = && \left(\frac{c_s l \rho_{0}}{\mu\mathcal{A}^{2}} - m_{\text{eff}}\frac{c_s}{\mu l^{2}}\right)c_s^{2}dt^{2} + 2 \frac{c_s^2m_{\text{eff}}}{\mu l^{2}}\left(1 - \sigma^{2} + \sigma \right)dt dr + 2 m_{\text{eff}} a_{\text{eff}} \left(\frac{\mu\mathcal{A}\sigma^{2}}{c_s l \rho_{0}}\right)\left(1 - \sigma^{2}\right) dt d\phi\nonumber\\
	&&+\left[\frac{c_s l \rho_{0}}{\mu\mathcal{A}}\left(1-\mathcal{A}+2\sigma\right)-m_{\text{eff}}\frac{c_s}{\mu l^{2}} \right]dr^{2} + 2 a_{\text{eff}} \sigma^2\left[\frac{1}{c_{s}\mathcal{A}}\left\{1-m_{\text{eff}}\frac{\mathcal{A}^{2}}{l^{3}\rho_{0}}\right\}+\frac{\sigma}{c_s} \right]dr d\phi\nonumber\\
	&&+\left[\frac{\mu \sigma^{4}}{c_s^{3} l \rho_{0}}a_{\text{eff}}^{2}\left(1-m_{\text{eff}}\frac{\mathcal{A}^{2}}{l^{3}\rho_{0}}\right)  -\frac{c_s l \rho_{0}}{\mu}r^{2}\right]d\phi^{2}~.\label{final_metric_in_terms_m_a}
\end{eqnarray}
\end{widetext}
The above acoustic metric is our final result. It is intriguing to note that our purely algebraic procedure of applying the coordinate transformation (\ref{generalised_transformation}) and somewhat heuristic prescription of identification of fluid mass and spin parameters have resulted in the acoustic metric (\ref{final_metric_in_terms_m_a}) which can be compared term by term with the Kerr metric (\ref{Kerr_metric_at_pi/2}). It is at once evident that $m_{eff},~a_{eff}$ in the acoustic metric (\ref{final_metric_in_terms_m_a}) occupy identical positions as $m,~a$ in the Kerr metric (\ref{Kerr_metric_at_pi/2}). This provides a mathematical consistency of our framework, but at the same time, the result calls for a deeper understanding of the physics behind the Analogue Kerr metric.

\section{Conclusion}
\noindent
In the present paper, we have considered sonic wave propagation in an extended fluid model with Berry curvature contributions in first-order perturbation. We have applied a number of quite strong restrictions on the fluid model, primarily for computational convenience. We have found that under a coordinate transformation, in polar coordinates with $\theta =\pi/2$, the acoustic metric developed in this model closely resembles the Kerr metric for spinning black hole in Eddington-Finkelstein coordinates also at $\theta =\pi/2$. Exploiting this mapping, we have provided explicit expressions for the effective mass and spin parameters of the acoustic metric. The present acoustic Kerr metric and subsequent derivation of effective Analogue Gravity parameters open up new possibilities.

As for future work, we intend to carry out analysis along the following directions:
\begin{enumerate}[label=(\roman*)]
    \item To study the wave dynamics and geodesic motion in this Analogue Gravity scenario. For the latter case, recent studies of \cite{Dalui2022} have shown a chaotic behaviour of particle motion near the Kerr Black Hole horizon. It might be worthwhile to look for similar effects in the acoustic black Hole case studied here.
    \item A deeper understanding of the coordinate transformation, performed in the present work, as well as the resulting mapping between the acoustic system  and Kerr model.
    \item In \cite{met}, one can find realistic and laboratory-based examples of physical systems that support a hydrodynamic flow of electrons. It will be worthwhile to look for signatures of the Analogue Kerr model in such systems. 
\end{enumerate}
\noindent
Finally, in this scenario where researchers are attempting to identify various  physical systems that can support analogue gravity situations that can be attained experimentally in a laboratory setup, we hope our work can shed some light in that direction.

\vskip 1cm

\begingroup
\renewcommand{\section}[2]{}

\endgroup

\end{document}